\documentclass[aps,prb,twocolumn,floatfix,superscriptaddress]{revtex4-1}
\usepackage{amsmath}

\usepackage{graphicx}
\usepackage{epsfig}
\usepackage{epstopdf}
\usepackage{grffile}

\usepackage{subfigure}
\usepackage{enumerate}
\usepackage{morefloats}

\usepackage[usenames,dvipsnames,svgnames,table]{xcolor}
\usepackage{hyperref} 
\hypersetup{backref,pdfpagemode=FullScreen,colorlinks=true,linkcolor=NavyBlue,citecolor=BrickRed,urlcolor=NavyBlue}

\usepackage{color}

\begin{document}
\newcommand{\rat}{$D_N/D_W=0.5$}
\newcommand{\ddu}{$D_N/D_W$}
\newcommand{\hund}{$J=U/4$}
\newcommand{\hyb}{$V/D=0.1$}
\newcommand{\chiloc}{$\chi_{loc}$}
\newcommand{\uosp}{$U_{OSP}$}
\newcommand{\mzd}{$\langle m_{z1} m_{z2}\rangle$}
\newcommand{\akw}{$A(\epsilon_k,\omega)$}
\newcommand{\gf}{Green's function}
\newcommand{\fl}{Fermi-liquid}
\newcommand{\os}{orbital-selective}
\newcommand{\hm}{Hubbard model}
\newcommand{\czz}{$C_{WN}$}

\title{Hybridizing localized and itinerant electrons: a recipe for pseudogaps}
\author{E. A. Winograd}
\affiliation{Laboratoire de Physique des Solides, UMR8502 CNRS-Universit\'e Paris-Sud, Orsay, France}
\author{L. de'~Medici}
\affiliation{Laboratoire de Physique et Etude des Mat\'eriaux, UMR8213 CNRS/ESPCI/UPMC, Paris, France}
\affiliation{Laboratoire de Physique des Solides, UMR8502 CNRS-Universit\'e Paris-Sud, Orsay, France}

\date{\today}

\begin{abstract}
In a system where selective Mott localization is realized, some electrons show a gap to charge excitations while others do not. A hybridization between these two kind of electrons will lead to a smoothening of this sharp difference and can even bring the system back to a complete delocalization. We show here that there is a large region of parameters at finite hybridization where the selective localization persists and the system shows a partial filling of the selective gap with incoherent states, giving rise to a pseudogap.
This result is illustrated here in a two orbital Hubbard model with Hund's coupling, but is based on quite general assumptions and should hold for a larger class of systems, and possibly be a paradigm for the pseudogap mechanism in cuprates.
\end{abstract}

\maketitle

\section{Introduction}

Mott insulators are a widely studied class of compounds. The localization of conduction electrons induced by their mutual Coulomb repulsion and the consequent impeded conduction are a striking manifestation of many-body physics. Indeed these materials are predicted metallic by one-body techniques (such as density functional theory) and it is only when dynamical correlations are taken into account that the observed insulating behavior can be accounted for. Typical examples of these materials are found in organic superconductors such as the Bechgaard salts and the Fullerenes, and in transition metal oxides like NiO, V$_2$O$_3$ and the high-Tc superconducting cuprates\cite{imada_mit_review}.

An intriguing spin-off of this research is the study of systems where orbital-selective Mott (OSM) localization happens\cite{vojta_OSMT_heavy_fermions_review_jltp_2010}. In these systems a subset of the conduction electrons undergoes Mott localization, while the rest remains itinerant and the system shows thus metallic conduction albeit in the presence of well-formed local magnetic moments. 

This physics is also the implicit starting point for understanding some old-standing problems like moment formation in some f-electron systems (like elemental Lanthanides and late Actinides) where electrons in incomplete f-shells remain localized while spd electrons are itinerant, but also in manganites La$_{1-x}$Sr$_x$MnO${}_3$, where electrons in the $t_{2g}$ subshell of the Mn 3d orbital manifold are localized and give rise to a strong local magnetic moment while the $e_g$ delocalize in conduction bands\cite{Salamon_Manganites_RMP}.

More subtly this selective localization can happen for electrons in the same subshell as it has been suggested for Ca$_x$Sr$_{2-x}$RuO$_4$\cite{anisimov}, LiV$_2$O$_4$\cite{Nekrasov_LiV2O4}, among others, and for the $\alpha$ phase of elemental iron\cite{Katanin_OSMT_alphaFe} and the iron-based superconductors\cite{luca09, demedici_Genesis}. This proposal has recently received a great deal of attention (see Ref.~\onlinecite{lucanew} for a list of references).

The two parts of the fluid are indeed not completely separated and two main effects couple them: hybridization and the electronic coulomb interaction.
Hybridization, i.e. the amplitude for an electron to jump between the more localized and more itinerant orbitals, is the dominant effect in some f-electron systems. Then the local moment can be quenched by quantum fluctuations and a metallic state with electrons bearing a strongly enhanced effective mass arises at low temperatures. These materials are known as heavy-fermions\cite{hewson}.
 
\begin{figure}
\centering
\includegraphics[width=0.9\columnwidth,angle=0]{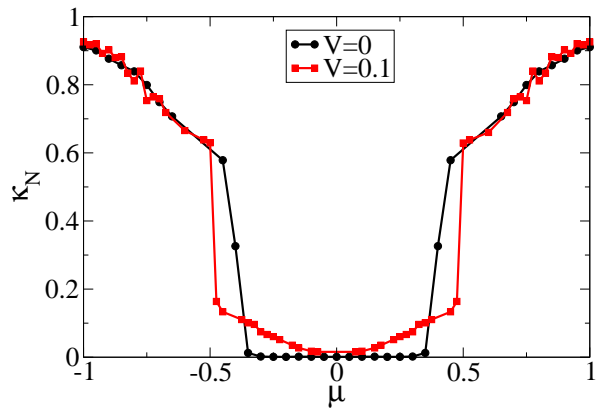}
\caption{\label{fig:comp2}Electronic compressibility of electrons in the localized orbital  ($U/D=1.4$, \ddu$=0.5$, $J=U/4$), calculated by numerical differentiation of $n(\mu)$ of Fig. \ref{fig:dens2}. 
The incompressible fluid found in absence of hybridization ($V=0$) due to a clean gap in the orbitally-resolved spectrum, is replaced at nonzero V by a phase in which the compressibility is reduced but non-vanishing. Excitations filling the gap are incoherent, and the pseudogap remains clearly delineated.}
\end{figure}

For d-only systems instead, the main coupling to be taken into account is the on-site coulomb repulsion between conduction electrons, and Hund's coupling - the exchange energy favoring the distribution of electrons in the different orbitals of the same shell with their spin aligned - plays a key role in these systems (see Ref. \onlinecite{Georges_annrev} for a recent review). This exchange term (which is also a measure of the reduced inter-orbital repulsion that electron feel when on the same atom, compared to the intra-orbital one) is known to favor a strong differentiation of the correlation strength among the different orbitals, ultimately leading to the selective Mott localization\cite{demedici_MottHund}.
 
Another effect due to the Hund's coupling is the overall reduction of the metallic coherence scale, whenever the sub-shell mainly responsible for the electronic states at the Fermi level is filled by more than one electron or one hole per site\cite{pruschke05, janus}.

In the realistic multi-orbital bandstructures of d-orbital materials, such as transition-metal oxides, inter-orbital hopping is generally present and non-local, since hops between orbitals of a given site are typically forbidden by symmetry. However structural distortions lowering the symmetry (like those introduced by substituting Calcium to Strontium in Ca$_x$Sr$_{2-x}$RuO$_4$) can introduce such hybridization terms.
All in all in systems where selective localization is expected on theoretical grounds without local and non-local hybridizations between the orbitals, one is led to ask if it will be spoiled by the presence of these terms in realistic situations and a fully metallic - possibly heavy-fermionic - behaviour will be recovered, as it happens in f-electron systems.

This issue is studied here on a minimal model of two half-filled bands of different bandwidth, with electrons interacting by means of local multi-orbital Coulomb repulsion (in the standard Kanamori form for Hund's coupling, see below) and in presence of local hybridization between the orbitals.  Non-local hybridization is very briefly touched upon at the end of the paper, with preliminary results confirming our main analysis.

The model is solved within Dynamical Mean-Field Theory (DMFT)\cite{bible}
that it is able to describe both itinerant and localized phases on the same footing. This method is formulated in the language of Green's functions, and the local excitation spectrum can thus be readily accessed.
In the single band case the Mott transition is signaled by the vanishing of the spectral weight at the Fermi level, present in the metallic phase,  with the spectrum opening a gap to charge excitations in the insulating phase. Correspondingly the population ($n$) ceases to grow with the chemical potential ($\mu$) when the latter lies within the gap, and the system becomes incompressible (i.e. $\kappa=\frac{dn}{d\mu}=0$).

The same analysis can be applied to the orbitally selective Mott transition which is the focus here, and the main result of this paper (which will be discussed more thoroughly in section \ref{osp}) is plotted in Fig. \ref{fig:comp2}. 

In absence of any hybridization between the two orbitals in the selectively localized phase the system has a gap in the spectrum for the creation of a charge excitation in the localized orbital,  while it has finite spectral weight at low energy for charge excitations in the delocalized one. The state is overall metallic and compressible but as function of the chemical potential the electron density in the delocalized orbital grows, while that in the localized orbital remains fixed to unity. One can thus say that within the electronic fluid there is an incompressible, insulating part. 

Upon the onset of hybridization one could expect metallicity to be restored in the latter, the gap to be filled by coherent excitations and that the electron fluid becomes wholly compressible again. What we have found is that under quite general assumptions the gap is only poorly filled and maintains clear edges,  thus forming a pseudogap.
The spectrum for the formerly localized orbital becomes filled by incoherent excitations and the incompressibility is replaced by a finite but very reduced compressibility.  

Thus, for all practical purposes the selective localization survives the onset of hybridization, be it local or non-local.
Our numerical DMFT calculations are performed at zero temperature, but have a finite energy resolution. They cannot distinguish if the excitations filling the pseudogap will acquire coherence below an energy scale that would be smaller than our resolution, or if this phase is genuinely non Fermi-liquid. This is discussed in section \ref{sec:discussion}. However the physical message is that at all temperatures relevant for experiments the selectively localized phase is robust to hybridization in a large region of the parameter space and that a depression in the physical compressibility should be clearly observable in experiments.

This issue was partially treated in previous works\cite{koga05, luca05}. Our results are compatible with these previous investigations and complete them (in Ref.~\onlinecite{koga05} only a few parameter values where explored, and in  Ref.~\onlinecite{luca05} slave-spin mean-field was used which cannot account for incoherent excitations) and give a complete phase diagram of the model, which is quite similar to the one expected on general grounds \cite{koga05}.

Finally it is worth mentioning that the present work may have a bearing on the issue of the formation of a pseudogap in the {\it k-space resolved} excitation spectrum in cuprates. Indeed a pseudogap is observed in a large region of temperatures and dopings around the directions $(0,\pi)$ of the Brillouin zone. This physics can be described in terms of a k-space-selective localization\cite{Venturini_MIT_cuprates, Biermann_nfl, Ferrero_dimer, gull10}. The hamiltonian describing the short-ranged correlations responsible for the pseudogap opening in the cluster extensions of DMFT that correctly capture this phenomenon and many other major features of the physics of cuprates, can be recast in a multi-orbital form that bears a strong similarity to the problem we solve in this paper (see the extended comparison in Ref. \onlinecite{lucanew}). Interaction terms acting in first approximation as hybridizations are present and possibly responsible of a smoothening of the gap due to the selective localization.

The paper is organized as follows: In section \ref{sec:modelandmethod} we describe the two-orbital model we analyze and the DMFT equations used to solve it. In section \ref{sec:results} we report the results of our numerical calculations: in \ref{phasediagram} the low-temperature phase diagram of the half-filled system is analyzed, as a function of interaction strength and local hybridization, and in \ref{osp} the OSM phase is studied as a function of the chemical potential.
In \ref{sec:discussion} we discuss the physics of the OSM phase as far as the actual ground state - and thus the ultimate fate of the system at strictly zero temperature - is concerned, and in section \ref{sec:conclusions} we trace our conclusions and perspectives.

\section{Model and method}
\label{sec:modelandmethod}

We explore this problem considering here a local hybridization. The case of non-local hybridization yields similar physics and is briefly mentioned in section \ref{sec:conclusions}. 
We consider the following two-band (describing a wide band and a narrow one, hence the subscripts W, N) Hamiltonian with on-site interactions, and local hybridization:

\begin{eqnarray}
\label{mbhm}
H&=&-\sum_{\substack{\langle i,j\rangle,\sigma\\l=N,W}}t_lc^\dag_{il\sigma}c_{jl\sigma}+
V\sum_{i,\sigma}\left(c^\dag_{iW\sigma}c_{iN\sigma}+
c^\dag_{iN\sigma}c_{iW\sigma}\right)\nonumber \\ 
&&-\sum_{i,\sigma,l}\mu c^\dag_{il\sigma}c_{il\sigma}
+H_{int}
\end{eqnarray}
where $c^{\dag}_{i,l,\sigma}$ and $c^{\phantom{\dag}}_{i,l,\sigma}$ are the fermionic creation and annihilation operators acting on site $i$, orbital $l$ 
and spin $\sigma$, $t_l$ is the orbital-dependent hopping amplitude and $V$ is the amplitude of the local hybridization. $H_{\textrm{int}}$ describes the multi-orbital electron-electron interaction and we use here the rotationally invariant Kanamori Hamiltonian, that reads
\begin{eqnarray}
\label{hintmbhm}
H_{int}&=&U\sum_{i,l=N,W}\tilde{n}_{il\uparrow}\tilde{n}_{il\downarrow}+U'\sum_{i,\sigma}\tilde{n}_{iW\sigma}\tilde{n}_{iN-\sigma}+\nonumber\\
&&+(U'-J)\sum_{i,\sigma}\tilde{n}_{iW\sigma}\tilde{n}_{iN\sigma}\nonumber-\\
&&-J\sum_i\left[
c^\dag_{iW\uparrow}c_{iW\downarrow}c^\dag_{iN\downarrow}c_{iN\uparrow}+
c^\dag_{iW\downarrow}c_{iW\uparrow}c^\dag_{iN\uparrow}c_{iN\downarrow}\right]-\nonumber\\
&&-J\sum_i\left[
c^\dag_{iW\uparrow}c^\dag_{iW\downarrow}c_{iN\uparrow}c_{iN\downarrow}+
c^\dag_{iN\uparrow}c^\dag_{iN\downarrow}c_{iW\uparrow}c_{iW\downarrow}
\right]
\end{eqnarray}
where for convenience, we write $\tilde{n}_{il\sigma}=n_{il\sigma}-1/2$, with $n_{il\sigma}$ the number operator on site $i$, orbital $l$ and spin 
$\sigma$. By doing this shift, which only adds constant and quadratic terms to the Hamiltonian, we guarantee that the particle-hole symmetric solution 
corresponds to $\mu=0$, that can be easily proven by performing a particle-hole transformation.
U is the strength of the intra-orbital Coulomb repulsion between electrons, $U'$ ($=U-2J$ for rotational invariance) is the inter-orbital one, and J is the exchange Hund's coupling.

It is important to point out that albeit it is possible to diagonalize the local part of the Hamiltonian
and eliminate the hybridization in favor of a crystal field splitting, in doing this, a non-local hybridization term 
appears (as long as $t_W\neq t_N$), so that the two transformed orbitals keep a finite tunneling amplitude between them. Hence, in general the hybridization introduces a new intrinsic complexity into the problem.

We solve this problem using dynamical mean-field theory\cite{bible}, where the lattice model is mapped into a single impurity problem obeying a self-consistency condition.
This allows to calculate non-perturbatively - through the numerical solution of the impurity problem, in general - the self-energy (which is then purely local) and other local correlation functions.

For simplicity we choose here customarily a semicircular density of states (which corresponds to the Bethe lattice), which simplifies the self-consistency condition and the results of which are usually general enough. 
We set the half-bandwidth of the widest band $D_W=D$ as the unit of energy.

The imaginary-time action of the two-orbital auxiliary Anderson impurity problem to be solved reads:

\begin{eqnarray}
S &\!=&\! -\int_0^\beta d\tau \int_0^\beta d\tau' \sum_{l,l',\sigma} c^\dag_{l\sigma}(\tau)\mathcal{G}_{0,ll',\sigma}^{-1}(\tau-\tau')c_{l'\sigma}(\tau ')\nonumber\\
 &\!+&\! \int_0^\beta d\tau H_{int} [\{c^\dag_{l\sigma}(\tau),c_{l\sigma}(\tau)\}]
\end{eqnarray}
where electron creation and annihilation operators here act on the impurity orbitals, $H_{int}$ is the local interaction of the original lattice problem and, in this case with local hybridization (in contrast to the special case of non-local hybridization that we mention in section \ref{sec:discussion}) the bare Green's function of the impurity problem $ \hat{\mathcal{G}}_{0,\sigma}(i\omega_n) $ is a non-diagonal $2\times 2$ matrix subject to the following self-consistency relation:
\begin{widetext}
\begin{equation}\label{eq:selfcons}
 \hat{\mathcal{G}}^{-1}_{0,\sigma}(i\omega_n) = \left(
  \begin{array}{cc}
   i\omega_n-t_W^2G_{WW,\sigma}(i\omega_n) & -V-t_Wt_NG_{WN,\sigma}(i\omega_n)\\
   -V-t_Wt_NG_{NW,\sigma}(i\omega_n) & i\omega_n-t_N^2G_{NN,\sigma}(i\omega_n) 
	  \end{array} \right)
\end{equation}
\end{widetext}
where  $G_{ll',\sigma}(i\omega_n)=-\int_0^\beta d\tau e^{i\omega_n\tau}\langle c_{l\sigma}(\tau)c^\dagger_{l'\sigma}(0)\rangle$ is the impurity Green's function (coinciding with the local Green's function of the lattice problem in the present framework), and $\omega_n$ are the fermionic Matsubara frequencies.

We solve the auxiliary impurity problem at zero temperature through the exact diagonalization\cite{caffarel,liebsch2012} (making use of the Lanczos algorithm) of the discretized problem in which the baths are expressed through finite number $N_s$ of non-interacting sites.
The self-consistency condition is enforced on a mesh of Matsubara frequencies $\omega_n=\pi(2n+1)/\beta$, where $D/\beta$ is the scale setting the energy resolution. The latter cannot be increased at will but has a lower limit, which is a function of $N_s$.
Standard calculations were performed at $\beta D=100$ with $N_s=10$, which means that we considered one site for each orbital plus eight for the bath. The results we present here were successfully benchmarked against analogous calculations with $\beta D=150$ and $N_s=11$ and $N_s=12$, where the low energy resolution is increased.

\section{Results}
\label{sec:results}
\subsection{The phase diagram at half-filling}
\label{phasediagram}

\begin{figure}
\centering
\includegraphics[width=\columnwidth,angle=0]{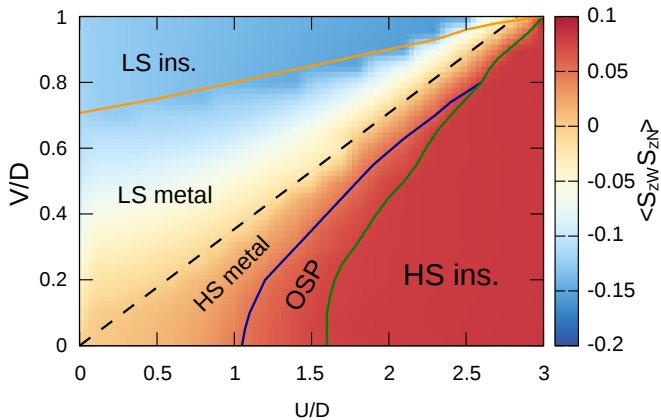}
\caption{\label{fig:phasediagramV}$V-U$ phase diagram of the half-filled 2-band \hm\ with \ddu$=0.5$ and $J=U/4$. The solid lines correspond to the metal-insulator boundaries. 
For weak coupling and hybridization, the system is metallic. For high $V$ and $U$, the system is band insulating with a low-spin $S=0$, and Mott insulator with a high-spin $S=1$, respectively.
For intermediate coupling and weak-to-intermediate hybridization, there is an \os\ phase, where only the narrow band localizes, even at finite hybridization. 
The color scale represents the inter-orbital spin-spin correlation function \czz, which is negative for the singlet state, and positive for the triplet. The metallic phase has a crossover from the limit in which the screened local moment is very small (labeled LS metal) to the one in which it is very close to the $S=1$ value favored by Hund's coupling (HS metal). The dashed line indicates the corresponding low-to-high-spin transition that occurs in the atomic limit, which is a good estimator 
of the crossover region.}
\end{figure}

We start by analyzing the general structure of the low-temperature phase diagram in the parameters $V-U$, which is plotted in Fig. \ref{fig:phasediagramV}.
We here focus on the metal-insulator boundaries which are represented by the solid lines and we will discuss the color scale, which represents the magnetic correlation functions, at the end of this section. Throughout the paper we set the narrow bandwidth $D_N/D_W=1/2$  and Hund's coupling $J/U=1/4$ (a brief mention of another set of values, $D_N/D_W=0.15$, $J/U=0.1$ yielding analogous results is done in section \ref{sec:discussion}). We start studying the particle-hole symmetric (half-filled) case $\mu=0$.

In absence of interactions (i.e. $U=J=0$), at zero hybridization the systems is metallic with two half-filled bands. The onset of the hybridization $V$ produces band-repulsion, giving rise to the bonding and antibonding bands. As $V$  increases, the bonding band gets more populated than the antibonding band, and a band-gap eventually opens when $V>\sqrt{D_ND_W}$ ($=\sqrt{0.5}$ in the present case). It is readily shown\cite{luca05} that upon onset of the interaction this criterion becomes:
 \begin{equation}\label{eq:bandgap}
 \left|V+\Sigma_{WN}(0)\right|>\sqrt{D_WD_N},
 \end{equation}
 where $\Sigma_{WN}(0)$ is the off-diagonal component of the local self-energy at zero frequency. In the weak-coupling limit it can be shown that $\Sigma_{WN}(0)\sim U-5J$ so that for the value of $J$ chosen in this study ($J=U/4$) the frontier moves to larger V upon onset of the interaction.

On the other hand at $V=0$, when increasing $U$ from the non-interacting limit, the system goes through three different phases. It is metallic at weak coupling and  Mott insulating at strong coupling as expected, whereas in an intermediate $U$-regime, an \os\ Mott phase appears where the Mott-gap opens only in the narrow band.\cite{anisimov, koga04,ferrero05,luca05,arita05,inaba07}

It is not a priori evident what happens to the OSM phase for finite $V$. 
Previous studies (focusing on a limited set of parameters\cite{koga05} or using slave-variable techniques\cite{luca05,lechermann07}) showed that the \os\ Mott phase seems to be unstable to a finite hybridization. 
We here explore the whole phase diagram as a function of U and V within DMFT which is the state-of-the-art technique.

\begin{figure}
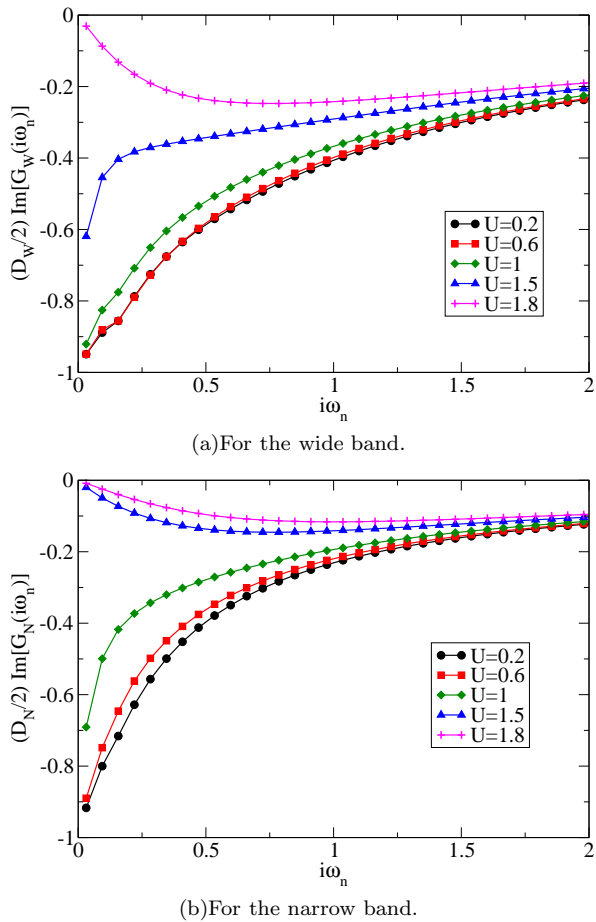

\centering
 \subfigure[For the wide band.]{\label{fig:Gv0.15w}
\includegraphics[width=0.9\columnwidth,angle=0]{fig3a.eps}}
 \subfigure[For the narrow band.]{\label{fig:Gv0.15n}
\includegraphics[width=0.9\columnwidth,angle=0]{fig3b.eps}}
\caption{\label{fig:Gv0.15} Imaginary part of the Green's functions for $V=0.15$, for different values of the interactions. 
Notice that for $U/D=1.5$ the system is in an \os\ Mott phase.}
\end{figure}

In Fig. \ref{fig:Gv0.15} we plot the Matsubara \gf{s} of both orbitals for different values of $U$ at $V=0.15$. We recall that extrapolating the \gf{s} to $\omega_n\to0$, a finite intercept indicates 
that the solution is metallic, while if the intercept vanishes, the solution is insulating. The \gf{s} are rescaled by the their corresponding bandwidth, 
which is their natural energy scale.

From this figure, it is clear that the situation is unchanged compared to the $V=0$ case. 
From weak to intermediate coupling, both bands behave metallic. Increasing $U$ further, there is a regime where the wide 
band behaves metallic and the narrow band is insulating. This is clearly shown by the \gf{s} at $U/D=1.5$. Then, for stronger interactions, both bands are insulating.

We trace in Fig. \ref{fig:phasediagramV} the border where we find the situation just depicted, and we see that there is a sizable zone in the U-V plane where the OSM phase persists even for finite V. This zone shrinks for increasing V and ultimately closes. Also, both boundaries for the OSM transition and the Mott metal-insulator one shift towards higher interaction strengths. 

Both these effects of V are easily rationalized when one realizes that as a first approximation V acts against Hund's coupling J and it simply counteracts its main effects. On the atomic spectrum the Mott gap  (as long as $J>V/\sqrt{2}$) for the half-filled case reads $\Delta_{\rm at}=U+J-2V$. Thus  J enlarges it and this has influence on the Mott transition happening in the lattice model\cite{demedici_MottHund}, with a strong reduction of the critical interaction strength $U_c$. 
V instead reduces this atomic Mott gap and then raises the $U_c$ in the lattice model back towards the values that it has at small J.

As we mentioned in the introduction, Hund's coupling also favors the orbital differentiation. It does so by quenching the local inter-orbital fluctuations\cite{koga05, luca09,demedici_MottHund}. Again, the hybridization contrasts this effect by favoring the low-spin high-orbital angular momentum atomic states, and thus enhancing orbital fluctuations. One can expect then that its effect will be to reduce orbital differentiation, as it is found in this study.

The metallic phase in which electrons in both orbitals are delocalized (i.e. excluding the OSM phase) is a Fermi-liquid. 
In this framework, quasiparticles have an associated weight - measuring metallicity - which in the single-band case reads 
$Z^{-1}=[1-\left.\partial\Sigma(i\omega_n)/\partial(i\omega_n)\right|_{\omega_n=0}]$. 
In the present multiband case we define the orbitally-resolved quasiparticle weight (i.e. $\Sigma$ and $Z$ are matrices) as $Z_l=[1-\left.\partial\Sigma_l(i\omega_n)/\partial(i\omega_n)\right|_{\omega_n=0}]^{-1}$. In the non-interacting case, 
$Z_l=1$ while it vanishes approaching the Mott state. 

\begin{figure}
\centering
\includegraphics[width=0.9\columnwidth,angle=0]{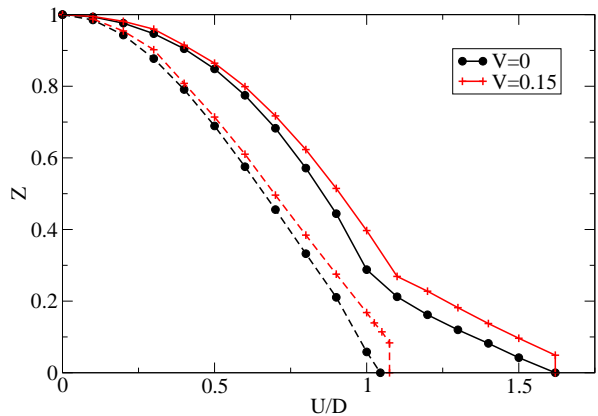}
\caption{\label{fig:qpv}Quasiparticle residue as a function of $U$. Solid (dashed) lines correspond to the wide (narrow) orbital.}
\end{figure}

In Fig. \ref{fig:qpv}, we plot the weight of the quasiparticle peak as a function of $U$ for each of the bands, for $V=0$ and $V=0.15$. 
Without hybridization, $Z$ continuously decreases until eventually vanishing for both orbitals. However, for finite $V$, the system enters the OSM phase (according to their corresponding Green's functions), when $Z$ is still finite for both orbitals (we signal this by plotting a ``jump" to a zero value for $Z_l$ in Fig. \ref{fig:qpv}, to signal the insulating behavior for orbital $l$).
The OSM phase is also metallic, but it has been shown to be non Fermi-liquid\cite{Biermann_nfl}, hence the above-defined quasiparticle weight loses its original meaning. We plot it nevertheless as it still gives a feeling of the robustness of the metallic phase. It is seen to still decrease and again not to vanish continuously when approaching the Mott insulating phase, as it happens instead for $V=0$.

It is important to remark that the OSM phase just mentioned is a physical phenomenon and not an artifact of the analysis of the Green's functions in the chosen basis\cite{demedici_Cerium}. Indeed at the OSM transition $\Sigma_{NN}(0)$ diverges, thus the determinant of the inverse Green's function, 
which is invariant under rotations, diverges as well. This means that at least one of the zero-frequency eigenvalues of the Green's function will vanish, which is what we define as an OSM phase, since it shows that there is a specific combination of electronic excitations without coherence\footnote{This means that in a ``wrong'' basis the OSM could go undetected if looking separately at the components of the Green's function only. However if detected in a specific basis (as we do here) it is certainly realized.}.
The trace of the Green's function at zero frequency (i.e. the spectral weight at the Fermi level) however, another basis-invariant, is finite in this phase, proving it metallic.
The OSM phase is thus a genuine physical phase, well distinct from both a normal metal and a Mott insulator.
This reflects in physical observables, which are indeed also basis-independent, such as the magnetic correlation functions to be discussed
later.

These results show sharp changes at the transitions which seem to suggest that both the OSM transition and the metal-Mott insulator one are first-order for finite $V$.

The order of the metal-insulator transition has been studied for the case $V=0$ in Refs. \onlinecite{pruschke05,ono}. These preliminary studies showed that it may be first-order for $J\not=0$, even at $T=0$. This issue is beyond the scope of the present work, and hence we did not study in detail the possible coexistence of both metallic and insulating solutions, which would be a signature of a first-order phase transition. However our results are consistent with a first-order phase transition.

However another explanation seems compatible with these results. We will discuss in Sec. \ref{sec:discussion} the nature of the ground state within the OSM phase at $V\not=0$.
We will point out in that section that at the frontier where the narrow band becomes incoherent there may simply be a sharp (in fact exponential) lowering of the Fermi-liquid coherence temperature, without real phase transition. However the crossing of this characteristic coherence scale below our numerical energy resolution may result in a sharp change in our results, as observed, without this implying an actual first-order phase transition at the metal-OSM phase frontier.

A metal-insulator transition happens also upon increasing of $V$ at finite U, as we mentioned before. The Green's functions behavior in this case is akin to the evolution of the non-interacting Green's functions, and the nature of the insulating state will be very similar to the band insulator found at $U=0$ and large V.

The present phase diagram bares remarkable similarities with the one for a system of two interacting bands with the same bandwidth and no hybridization but a crystal-field splitting of the local energies (explored in Ref. \onlinecite{hsls} and \onlinecite{Kunes_hi-lo}, and discussed in Ref. \onlinecite{Georges_annrev}), although the two systems differ for both the bandwidth ratio and for the fact that the hybridization cannot be eliminated in our system, as mentioned in section \ref{sec:modelandmethod}. 
Common points are the re-entrant shape of the metallic phase signaling compensation between the effects of Hund's coupling and hybridization/crystal field and the fact that the insulating phases are connected through a low-spin (LS) to high-spin (HS) transition.

Indeed a clear difference between the insulating states at large V/small J and at large J / small V  is expected because in the atomic limit there is a level crossing at $V=\sqrt2J$, and a corresponding  LS-HS transition from an $S=0$ to an $S=1$ ground state.
By inspecting the spin-spin correlation functions one can see how the formation of the local moment happens, progressively, in the lattice model.

The color scale of the phase diagram of Fig. \ref{fig:phasediagramV}, represents the value of the off-diagonal spin-spin correlation function 
\czz$=\langle S^z_W(0)S^z_N(0)\rangle$.
In the non-interacting case and without hybridization, \czz$=0$. As $V$ increases, a singlet state is favored,
and therefore \czz\ becomes negative. In the atomic limit, \czz$=-1/4$.
Upon increasing interactions, which implies an increasing $J$, the spin triplet state is favored instead, and \czz\ becomes positive and eventually saturates, when the system is deeply in the Mott state, to the value of the atomic limit for the $S=1$ triplet \czz$=1/12$.

In Fig. \ref{fig:phasediagramV} we also trace as a dashed line the boundary for the LS-HS transition ($V=\sqrt2J$) for the atomic limit, in which \czz\ abruptly changes from $-1/4$ to $1/12$.  
Notice that in the lattice model the transition is replaced by a smooth crossover, yet well represented by the atomic limit result.

\begin{figure}
\centering
 \subfigure[As a function of $U$ for $V=0.2$]{\label{fig:szv02}
\includegraphics[width=0.9\columnwidth,angle=0]{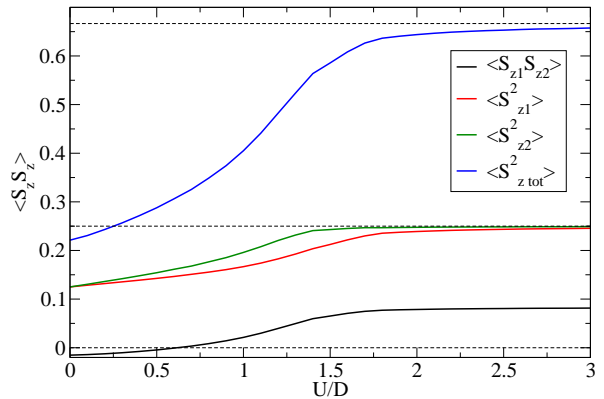}}
\subfigure[As a function of $V$, for $U=2.0$]{\label{fig:szv0}
\includegraphics[width=0.9\columnwidth,angle=0]{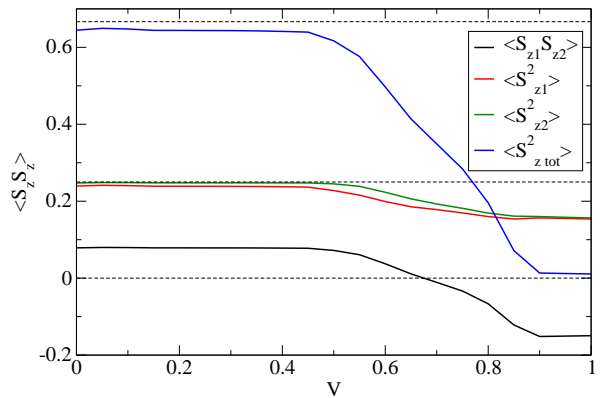}}
\caption{\label{fig:sz} Spin-spin correlation functions.}
\end{figure}

In Fig. \ref{fig:sz} we plot all the (z-axis) components of the orbitally resolved spin-spin instantaneous correlation function $\langle S^z_\alpha(0)S^z_\beta(0)\rangle$ ($\alpha,\beta\! =\! N,W$), and their sum $\langle S_{z\;\rm tot}^2(0)\rangle$ (with $S_{z\;\rm tot}=S_{zW}+S_{zN}$), in a typical case.

The $\langle S^2_{z\;\rm tot}\rangle$ illustrates the saturation to the atomic value for a spin-1 $\langle S^2_{z\;\rm tot}\rangle=2/3$ in the Mott insulating phase. Once decomposed one sees that the two orbitals also form an instantaneous local moment saturating each to the value for a spin-1/2 $\langle S^2_{z\;\rm tot}\rangle=1/4$. However they do this independently, owing to the existence of the OSM phase, in which one orbital has an atomic instantaneous local moment, while the moment in the other orbital is still reduced by quantum fluctuations. 

Regardless of the instantaneous local moment then, when a metallic phase is formed for both orbitals, the local moment is quenched and reduces to zero on a time scale inversely proportional to the Kondo temperature, in order to establish the Fermi-liquid coherence.
One expects however quite some difference in the Kondo scale depending on the size of the instantaneous local moment to be screened. Indeed screening a high-spin impurity results in an exponential reduction of the Kondo temperature, compared to a low-spin one (see discussion in Ref. \onlinecite{Georges_annrev} and references therein).
The system under examination shows a gradual crossover between the two extreme cases of screening a very small local moment (close to the spin-singlet state), named ``LS metal'', and screening the $S=1$ local moment induced by Hunds coupling named ``HS metal''. Thus very different coherence temperatures are expected in the different parts of the phase diagram.

Finally, in the OSM phase the instantaneous local moment is only partially screened. A residual spin-1/2 remains unscreened at long times, which acting as a scatterer prevents the Fermi-liquid behaviour from being realized. Indeed the bad metallic behavior\cite{Liebsch_Costi_nFL} following from this non Fermi-liquid (NFL) physics of the OSM phase has been characterized as having an anomalous self-energy similar to the one of the Ferromagnetic Kondo model, at low energy\cite{Biermann_nfl,greger2013}:
\begin{equation}\label{eq:nfl}
\textrm{Im}\Sigma^{\textrm{NFL}}_W(i\omega_n)=\frac{-a}{\ln^2(i\omega_n)+b\ln(i\omega_n)+c}.
\end{equation}

\begin{figure}
\centering
\includegraphics[width=0.9\columnwidth,angle=0]{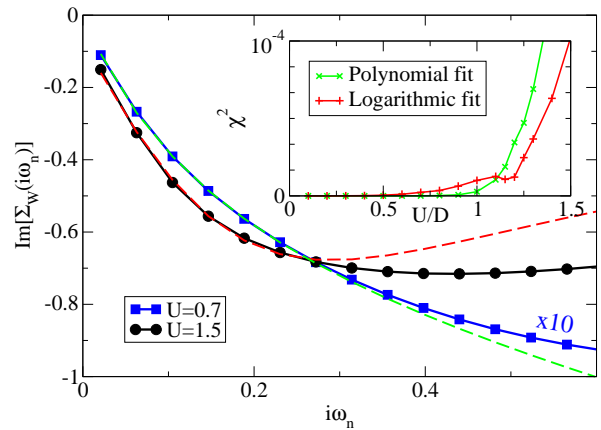}
\caption{\label{fig:sigmafit}Wide orbital self-energy in the metallic ($U/D=0.7$) and selective ($U/D=1.5$) phases (for $V/D=0.2$). Red dashed line shows the fit with the non Fermi-liquid $\Sigma$ of Eq. \ref{eq:nfl}, and the green dashed one corresponds to a polynomial fit. The inset shows that below the critical $U$ of the metallic-OSP transition, the polynomial fit, corresponding to the Fermi-liquid solution, is better than the logarithmic one, and the opposite behavior starts at the transition, suggesting that the metal is in a non Fermi-liquid state.}
\end{figure}

The self-energy of the itinerant component, in the OSM phase, is indeed well fit by Eq. \ref{eq:nfl}, as shown in Fig. \ref{fig:sigmafit}, where we also show a polynomial (i.e. Fermi-liquid, FL) fit of the self-energy for the case $U/D=0.7$ which lies inside the wholly metallic state. In the inset of Fig. \ref{fig:sigmafit}, we plot $\chi^2=\sum_n^{N_{\textrm{max}}}\left\{\textrm{Im}\left[\Sigma_W(i\omega_n)\right]-\textrm{Im}\left[\Sigma_W^\alpha(i\omega_n)\right]\right\}^2$, for $\alpha=$NFL and $\alpha=$FL. In both cases we used 3 parameters to fit and use the same number of fitting points ($N_{\textrm{max}}=6$). From this plot we can clearly see that starting from the weak coupling limit, the Fermi-liquid fit is better up to $U=U_\textrm{OSP}$, where the $\chi^2$ curves cross. Increasing $U$ further, the logarithmic fit of Eq. \ref{eq:nfl} becomes more representative of the metallic state, signaling the non Fermi-liquid state. We expect that the differences in the two $\chi^2$ would be even larger if we had access to lower energy scales.

\subsection{Doping the orbital-selective Mott phase: pseudogap behavior.}
\label{osp}

In this section we wish to characterize further the orbital-selective phase of the phase diagram of Fig. \ref{fig:phasediagramV}. 
Instead of plotting the spectral function, that within the ED is discretized, we explore the charge dependence on the chemical potential, that gives an approximate account of the excitation spectrum across the gap, yet it is smooth in our method.
 
\begin{figure}
\centering
 \subfigure[Total density, and density for each orbital]{
\includegraphics[width=0.9\columnwidth,angle=0]{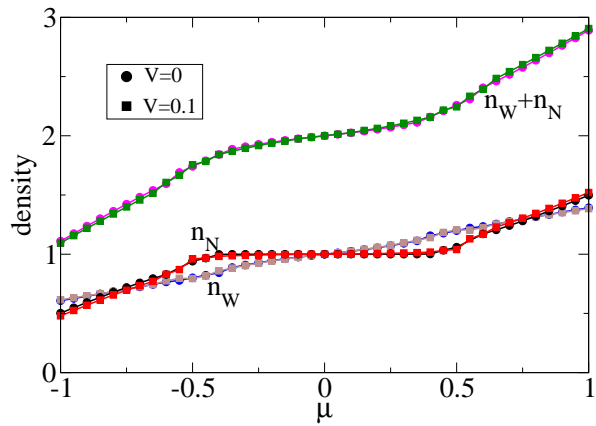}}
 \subfigure[Density of the narrow band.]{\label{fig:dens2}
\includegraphics[width=0.9\columnwidth,angle=0]{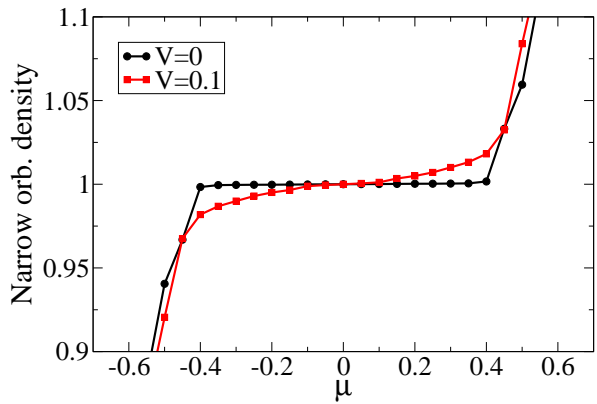}}
\caption{\label{fig:density} Charge density as a function of the chemical potential for $U/D=1.4$. Circles (squares) correspond 
to $V=0$ ($V=0.1$). The solutions for $\mu<0$ were obtained by a particle-hole symmetry.}
\end{figure}

The total and orbitally-resolved population is plotted in Fig. \ref{fig:density}, as a function of the chemical potential, 
for $U/D=1.4$ and both for $V=0$ and for $V/D=0.1$, values within the OSM phase of the phase diagram.

From Fig. \ref{fig:density}, we see that the system behaves metallic for both values of the hybridization, 
since this state is characterized by a monotonically increasing total number of particles as $\mu$ increases. From the same plot, we can also see that the same qualitative behavior occurs in the wide orbital, so the wide band is metallic.

The behavior of the narrow band is more subtle, and it is also plotted in Fig. \ref{fig:dens2} in a different scale. While for $V=0$ this band has a clear plateau at $n=1$, which signals the strong suppression of charge fluctuations in the incompressible Mott state 
($\kappa_N(\mu)=\partial n_N/\partial\mu=0$), in the hybridized case, the narrow band has also a strong reduction of charge fluctuations but no plateau develops. 
Hence, this band displays strong features of a Mott state, namely the reduction of the charge fluctuations below some critical $\mu_c(V)$ and its insulating behavior at $\mu=0$ according to the \gf, but, however, no gap opens. We identify this anomalous state as the opening of a pseudogap in the narrow band.

This pseudogap is clearly seen by looking at the compressibility on this band (which is simply the derivative of the curves in Fig. \ref{fig:dens2}), reported in Fig. \ref{fig:comp2}. 
Without hybridization there is a well-defined incompressible state, while for finite $V$ the compressibility gets strongly reduced without vanishing. Still, this drop is very sharp, and the pseudogap maintains clear edges.

Hence in both cases a clear signature of the OSM phase should be visible in the compressibility of the system. 

\section{Discussion: the actual ground state in the OSMP}
\label{sec:discussion}

In Sec. \ref{phasediagram}, we showed by analyzing the \gf{s}, that in the two-band Hubbard model under examination, with sizable Hund's coupling J, the narrowest band becomes insulating above a critical $U$, while the widest one remains metallic, thus realizing and OSM phase both in absence and presence of hybridization between the bands. Furthermore, the abrupt change in the quasiparticle weight of Fig. \ref{fig:qpv}, suggests that the metal-insulator transition in each band might be of first-order for finite $V$.

This problem was partially addressed in previous studies. 
Koga et al.\cite{koga05} showed that for a sizable value of $V$ the OSM phase is turned back into a fully metallic one with both orbitals delocalized, which is indeed compatible with our results. Relatively high temperature and a large mesh of values of V however prevent the use of data in Ref. \onlinecite{koga05} for deciding if the OSM phase is replaced by a heavy-fermionic behavior for any value of hybridization, however small. 
In a previous work\cite{luca05} of one of the authors and coworkers it was suggested that the heavy fermionic behavior would always replace the OSM phase at any finite $V$, but that the coherence temperature established would in many cases be exponentially reduced compared to the bare one. However the method used in Ref. \onlinecite{luca05} to address this issue is a slave-variable mean-field, which only captures correctly Fermi-liquid phases, thus its bearing on the realization of the incoherent, pseudogapped, non Fermi-liquid phase we study in the present work is questionable.  Such a phase is probably described within these approximation as a Fermi-liquid phase with a very small coherence scale, ultimately preventing the method from distinguishig between a pseudogapped and a heavy-fermionic phase.

Our present analysis indeed shows that a substantial region of parameters does exists where $V$ is not big enough to restore a fully coherent metallic phase.

However a word of caution is in order.
Despite DMFT can correctly capture NFL phases like the one we present in this article, it is important to remark that within the exact diagonalization method of solution of DMFT, that we use here, the resolution of small energy scales is finite, as mentioned in Sec. \ref{sec:modelandmethod}, and thus an extremely small Kondo scale can go undetected.

\begin{figure}
\centering
\includegraphics[width=0.9\columnwidth,angle=0]{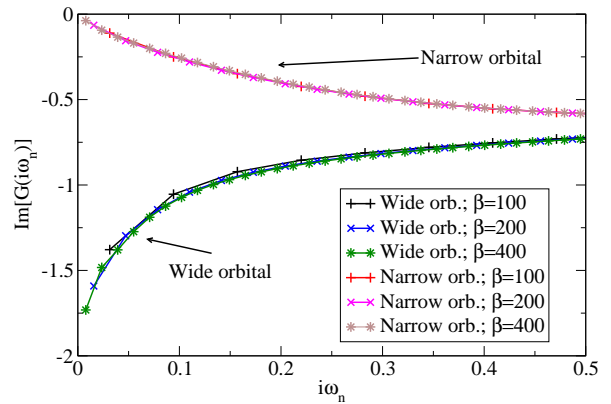}
\caption{\label{fig:Gosp}Imaginary part of the Green's functions for $U/D=1.5$ and $V/D=0.2$, for different values of the fictitious temperature $\beta^{-1}$.}
\end{figure}

Indeed the resolution is determined by the number of sites in the bath and from the mesh in Matsubara frequencies we use in the implementation of the self-consistency conditions of Eq. \ref{eq:selfcons}. The fictitious inverse temperature 
$\beta$ parametrizing this mesh cannot be brought to values that would uncover the discretized nature of the baths\cite{caffarel}\footnote{It is to be noted however that the adaptive nature of the self-consistent bath makes this method still very powerful compared to the resolution that one may expect from discretizing any continuum system. It is a standard performance to easily resolve scales of the order of 1/100 of the bare energies involved, as it is done in this work\cite{caffarel}}. By systematically enhancing this resolution to the best of our possibilities, we found these phases to be stable, which implies that the phase diagram of Fig. \ref{fig:phasediagramV} is accurate up to a very small energy scale, compared to the bare scales of the model ($D_W,D_N,U,J,V$). In Fig. \ref{fig:Gosp} we plot the \gf{s} in the \os\ phase, with different values of $\beta$, where we show  that when we increase the low energy resolution (increase $\beta$), the tendency of our results remains unchanged. We also considered 10 to 12  sites in the exact diagonalization calculations, and we obtain very similar results, with only a small discrepancy in the critical values. \footnote{The case with 11 sites (or any other odd number of $N_s$) is particularly important to this 
study, since one of the orbitals is forced to have an odd number of bath sites, which we chose to be the narrow one. This implies that, in the particle-hole symmetric case (as the case studied in Fig. \ref{fig:phasediagramV}), a pole of the narrow orbital Green's function is always located at $\omega=0$.
Hence there is always a finite low energy spectrum, which weight is only controlled by the degree
of hybridization between the `odd' bath-site and the orbital.
This configuration favors the metallic solution of this band, 
and thus the \os\ solution is harder to achieve. As seen with $N_s=10$ and $12$, even in this disadvantageous case we found the Mott \os\ phase to be robust.}

Despite all these considerations, this method does not let us assert definitely the absence of a smaller scale which can make the orbital-selective 
phase disappear as put forth in Ref. \onlinecite{luca05}. To this end, a numerical renormalization group calculation of this model would be appropriate.  

Anyway the solid result of our study is that a boundary in the space of parameters marks a sharp change of the metallic behavior for all practical purposes (meaning by this that if a conventional metal is restored in the place of the OSM phase this will probably happen outside of the temperature range of experimental interest). Hund's coupling overall reduction of coherence scales certainly enhances the effect. 

One may argue that for smaller Hund's coupling ratio $J/U$ this scenario is less likely, and it is indeed true that the OSM phase disappears below a critical ratio\cite{luca05,ferrero05,luca09}. However this critical ratio gets lower for more extreme bandwith ratios (and eventually one always has an OSM phase for $D_N/D_W\lesssim 0.2$). We have performed a similar analysis for a model with $D_N/D_W=0.15$ and $J/U=0.1$ and we find analogous results, with a large OSM phase both in absence and presence of local hybridization, and as robust to refinements of the numerical resolution as the one reported in the present work.

Regardless of the existence or not of a tiny Fermi-liquid scale below our resolution also in the OSM phase, an appealing description of the switching between the two metallic phases can be given in terms of the two screening channels of the local moment forming in the more correlated ``N" orbital\cite{demedici_Cerium}. Indeed there is a screening due to the electrons hopping from neighboring N orbitals, and a screening due to the electrons in the W orbital (this dichotomy can be tracked also in the magnetic susceptibilities\cite{Amaricci_doublescreening}). The former can be seen as of the ``Hubbard" type, meaning that the effective bath from which the electrons hop on the impurity in the DMFT description (which reads $t_N^2G_{NN}(\omega)$) has a density of states that is directly related to the Green function in the N orbital as in the DMFT description of the single-band Hubbard model. If $V$ is small this screening channel will prevail and as in the Hubbard model the self-consistent adaptation of the bath makes the Kondo energy scale (roughly the quasiparticle weight in our framework) linear, as we observe in the Fermi-liquid phase in Fig. \ref{fig:qpv}.
The second screening channel is of the "Periodic Anderson Model" (PAM) type (it reads, neglecting correlations involving the electrons in the W orbital, $[V+t_Nt_WG_{NW}(\omega)]^2/[\omega-t_W^2G_{WW}(\omega)]$) and, when the first channel becomes less effective with the increasing correlation strength, tends to take over. This marks the switching from a Hubbard-type physics to a PAM-type physics, in which the bath screens the impurity in a way similar to a non-selfconsistent impurity, i.e. the Kondo scale is exponential in both $U, V$.
In absence of Hund's coupling this switching would mean that the quasiparticle scale changes from the linear form to an exponential form, compatible also (in particular for small $V$) with a very quick crossing of our resolution and a consequent sudden loss of the Fermi-liquid behaviour, and thus with our results.
However the Hund's coupling adds further complexity to the problem. Indeed it may only lower further the Kondo scale, thus making the Fermi-liquid phase even more unreachable and thus favoring the OSMT.
However, as investigated in Ref. \onlinecite{Koga_Ext-PAM}, an NRG study in an extended Periodic Anderson model with Hund's coupling (i.e. our model with $t_W=0$, a sensible approximation of the present OSM phase) in presence of $J$ there may be a critical hybridization, due to the competition between Hund's ferromagnetic and the hybridization's antiferromagnetic Kondo couplings of the localized orbital with the itinerant electrons, below which the screening does not happen. If this is the case, the OSM phase may be a genuine zero-Temperature non Fermi-liquid phase. 

\begin{figure}
\centering
\includegraphics[width=0.9\columnwidth,angle=0]{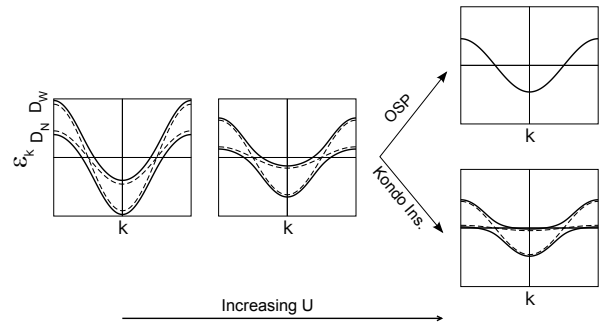}
\caption{\label{fig:conjecture} Possible zero-temperature phases in cases where coherence is restored at T=0. Solid lines correspond to the hybridized bands, while dotted lines correspond to the case $V=0$. As interactions increase, the bands get narrower and the effective hybridization decreases. When interactions are high enough, depending on the value of $\Sigma_{NW}(0)$ either (i) a Mott-like gap opens in the narrow band, renormalizing $V_{\rm eff}=0$, and only one band crosses the Fermi level; or (ii) a lower energy scale arises, opening a Kondo gap for any $V_{\rm eff}\not=0$.}
\end{figure}

Finally, if Kondo screening actually sets in at an extremely low temperature, two possible ground states are still possible, as suggested by Koga, et. al.\cite{koga05b}.
Indeed, as depicted in Fig. \ref{fig:conjecture},  either a heavy-fermionic metallic ground state is realized or a Kondo insulator, where a tiny band gap is opened between the coherent bonding and antibonding bands, owing to the integer filling. 

It is easy to show\cite{luca05} that the criterion for the opening of such a gap is Eq. \ref{eq:bandgap}, in which electronic interactions only enter through $\Sigma_{NW}(0)$. If $|V+\Sigma_{NW}(0)|$ is small enough (which most probably always happen at small enough $V$ since, for vanishing $V$, $\Sigma_{NW}(0)$ should be linear in $V$) a heavy metal is realized, otherwise a Kondo band gap is opened. 
We cannot decide which of the ground states is realized in our case because the extrapolation of $\Sigma_{NW}$ to zero frequency cannot be done from our numerical data, which are representative of the incoherent phase in any event.
If coherence is established below our resolution, one expect a change of behavior in all self-energies and thus numerical data below our resolution should be used to correctly extrapolate $\Sigma_{NW}(0)$.

\section{Conclusions}
\label{sec:conclusions}

In this paper we explored the effect of hybridization on the orbital-selective Mott phase, through a study of a two-band Hubbard model with different bandwidths $D_W, D_N$ for the two bands and local multi-orbital interactions with Hund's coupling.

We have explored the $U-V$ phase diagram for a typical choice $D_N/D_W=0.5, J=U/4$ for which the OSM phase is realized and have shown that this phase is stable, for all practical purposes, under small to intermediate hybridization (roughly as long as the local high-spin state is realized in this model, i.e. as long as $V\lesssim \sqrt{2} J$). 

Furthermore we have shown that the gap to charge excitations in the localized band typical of the OSM phase is turned into a pseudogap by the onset of hybridization, where incoherent excitations bring in spectral weight all the way down to zero energy. The pseudogap is clearly defined and a clear signature should be visible in the total compressibility of the system.

Generally speaking, the outlined ``recipe for pseudogap'', should be actually more general than the somewhat specific model that we explored. Indeed other studies\cite{demedici_Cerium, deLeo_PAM_CDMFT_PRB,deLeo_PAM_CDMFT_PRL} of hybridized localized and itinerant electrons have reported similar features in the spectrum. These studies together with ours seem to support a common scenario in which hybridizing localized and itinerant electrons can easily produce a metallic phase with a very low coherence temperature, as it happens in heavy-fermions. On top of this the presence of an exchange coupling (Hund's coupling in our case, RKKY coupling in Refs.\onlinecite{ deLeo_PAM_CDMFT_PRB,deLeo_PAM_CDMFT_PRL}) in general further reduces the coherence scale at an extremely low temperature and can even stabilize a genuine OSM phase at zero temperature in some range of parameters. When this happens we have shown that a clear signature in the spectrum and compressibility is found in the form of a pseudogap.

Experimentally a pseudogap has been found and thoroughly studied in underdoped cuprate superconductors.
Establishing a precise link between our simple model and the vast body of literature on the subject is beyond the scope of the present article, however we quickly outline here why our study may have a bearing on it. As mentioned in the introduction the pseudogap in cuprates has been recast in the language of selective Mott transition in momentum space\cite{Venturini_MIT_cuprates, Biermann_nfl, Ferrero_dimer, gull10}, meaning that the selective localization that we study here in orbital space, may instead happen for the electrons in the different parts of the Brillouin zone. In these studies - that use the single-band Hubbard model and capture all the main features of the low-temperature phase diagram of the cuprates - the electronic correlations responsible for this phenomenology are calculated through a multi-site (or equivalently multi-orbital) cluster model. The cluster hamiltonian has a form very similar to the standard Kanamori hamiltonian we use here. Supplementary terms have the form of scattering vertex which may be treated, as a first approximation, as hybridization terms between the orbitals. We suspect then that the rounding of the gap in k-resolved spectrum may be the result of our ``recipe'' when applied to the cluster hamiltonian, thus rounding the gap of the k-space selective Mott transition into the pseudogap observed in cuprates.

More concretely, a recent ARPES study\cite{Lin_Brouet-Pseudogap_FeTe} of Fe$_{1.06}$Te,  reported the appearance of a pseudogap on the electron pockets while the hole pockets have the expected intensity at the Fermi surface.
We deem this phenomenon a clear manifestation of the orbital selective coherence suggested to happen in iron-based superconductors (see Ref. \onlinecite{lucanew} for an experimental and theoretical overview). Among them FeTe is considered the most strongly correlated\cite{Yin_kinetic_frustration_allFeSC, lanata}, and where the OSM phase is most likely to be realized. Clearly the high-temperature (i.e. $T\gtrsim 80K$) paramagnetic phase is bad-metallic and shows selective coherence on the different Fermi sheets (besides the pseudogap on the electron pocket, the sheet mainly coming from the Iron $d_{xy}$ orbital, the more correlated in all studies, is not observed at all). This strong departure from Fermi-liquid transport characterized in Ref. \onlinecite{Biermann_nfl} and the pseudogap studied in this work are signatures expected in an OSM phase at finite temperature.  

Let us finally remark that the physics due to the hybridization that we report in this article may also occur in the presence of non-local hybridization (instead of the local $V$ considered here), as the one considered in Ref. \onlinecite{poteryaev08}. We performed some calculations on that model, and we obtained similar effects upon the onset of the non-local $V$. Hence, we deem that the pseudogap phase we report in this paper is not restricted to the particular model we studied, but is probably a general behavior coming from the competition between Hund's coupling and hybridization in the OSM phases.

The authors thank for fruitful discussions Veronique Brouet, Massimo Capone, Lorenzo De Leo, Marcello Civelli, Marcelo Rozenberg, and Michele Fabrizio particularly.
Both the authors were funded by CNRS and Agence Nationale de la Recherche under program ANR-09-RPDOC-019- 01.


%
\end{document}